\documentclass[conference]{IEEEtran}

\IEEEoverridecommandlockouts
\usepackage{cite}
\usepackage{amsmath,amssymb,amsfonts}
\usepackage{subcaption} 

\usepackage{algorithm}
\usepackage{algpseudocode}

\usepackage{graphicx}
\usepackage{textcomp}
\usepackage{xcolor}
\def\BibTeX{{\rm B\kern-.05em{\sc i\kern-.025em b}\kern-.08em
    T\kern-.1667em\lower.7ex\hbox{E}\kern-.125emX}}

\usepackage{graphicx} 
\usepackage{fancyhdr} 
\usepackage[utf8]{inputenc}
\usepackage{longtable} 
\usepackage{ragged2e}  
\usepackage{booktabs}  
\usepackage{caption}   
\usepackage{graphicx} 
\usepackage{float}    
\usepackage[utf8]{inputenc}

\usepackage{xcolor}

\usepackage{amsmath}        
\usepackage{graphicx}       

\usepackage{cite}           
\usepackage{xurl}           

\begin{document}

\title{

GPU Programming for AI Workflow Development on AWS SageMaker: An Instructional Approach\\


}

\author{
\IEEEauthorblockN{
Sriram Srinivasan, Hamdan Alabsi, Rand Obeidat, Nithisha Ponnala, Azene Zenebe
}
\IEEEauthorblockA{
Bowie State University, MD, USA \\
\{ssrinivasan, halabsi, robeidat, azenebe\}@bowiestate.edu, \\
ponnalan0103@students.bowiestate.edu
}
}
\maketitle

\begin{abstract}
We present the design, implementation, and comprehensive evaluation of a specialized course on GPU architecture, GPU programming, and how these are used for developing AI agents. This course is offered to undergraduate and graduate students during Fall 2024 and Spring 2025. 
The course began with foundational concepts in GPU/CPU hardware and parallel computing and progressed to develop RAG and optimizing them using GPUs. Students gained experience provisioning and configuring cloud-based GPU instances, implementing parallel algorithms, and deploying scalable AI solutions.
We evaluated learning outcomes through assessments, course evaluations, and anonymous surveys. The results reveal that (1) AWS served as an effective and economical platform for practical GPU programming, (2) experiential learning significantly enhanced technical proficiency and engagement, and (3) the course strengthened students’ problem-solving and critical thinking skills through tools such as TensorBoard and HPC profilers, which exposed performance bottlenecks and scaling issues. Our findings underscore the pedagogical value of integrating parallel computing into STEM education. We advocate for broader adoption of similar electives across STEM curricula to prepare students for the demands of modern, compute-intensive fields.

\end{abstract}

\begin{IEEEkeywords}
GPU, Parallel Computing, AWS SageMaker, HPC profiling, AI agents, Deep Learning
\end{IEEEkeywords}

\section{Introduction}
We are currently entering an era where Artificial Intelligence (AI) agents are capable of reasoning, planning, and acting with unprecedented autonomy. To build these sophisticated AI agents, High-Performance Computing (HPC), and specifically Graphics Processing Unit  (GPU) programming skills, play a crucial role. Traditional computing paradigms simply cannot handle the immense computational demands of modern AI. Training deep learning models and deploying real-time, high-fidelity AI solutions requires a level of parallel processing power that only GPUs can deliver. There is an exploding demand for AI engineers and data scientists \cite{nvidia_ai_demand}\cite{alooba_parallel_2024}\cite{bls_data_scientist_2024} who understand how GPUs work and how to exploit parallelism in GPU architecture to design intelligent algorithms. 
Foundational GPU programming skills are rapidly becoming an indispensable competency for anyone pursuing a career in AI. Consequently, infusing GPU programming into graduate and undergraduate STEM curricula is a critical necessity to prepare the next generation of scientists, engineers, and data professionals.

In Fall 2024 and Spring 2025, we taught an elective course (Special Topics). This course was taken by seniors majoring in Data Analytics and Information Systems who have completed an Introduction to Programming course. It's also offered as a cross-listed elective for graduate students in Information Systems. In both semesters, our focus was on developing AI workflows and agents with GPU programming using libraries like RAPIDS and Dask\cite{rocklin2015dask}. The first half of the 16-week course concentrated on introducing basic HPC concepts, demonstrating how to use accelerators to improve performance, and implementing simple parallel algorithms on accelerators. Previously, when this course was offered, we never introduced HPC concepts. Two factors contributed to revising the curriculum and infusing HPC for developing AI agents: 1) The support received through The U.S. National Science Foundation (NSF) funding, specifically the NSF Expand AI grant, which enabled us to procure Amazon Web Services (AWS) credits to facilitate this transition \cite{nsf2401658}. 2) The availability and accessibility of infrastructure provided by AWS, which offers affordable, on-demand access to GPU instances as a cloud service. This capability eliminated the need for maintaining on-campus HPC infrastructure to support the course.

Combining both semesters, we had about thirty-nine students enrolled in this course. All students had a Python programming background, as the prerequisite course (Introduction to Programming) was taught in Python. 
Most students lacked C programming knowledge, and since the course's focus was on developing AI agents using HPC, they all utilized Python JIT (Just-in-Time) libraries such as Numba\cite{numba_unified_memory2020} \cite{lessons_cuda_numba2020} \cite{numba_jit_ref} and CuPy\cite{mdpi_montecarlo2023} for parallel programming. As most of them had no prior knowledge of parallel computing, the course began with the basics of writing simple parallel programs and running them on GPUs. Students were given AWS console/Educate access during the first class, where they set up credentials and were taught how to use Python scripts to spin up and terminate instances. Students were familiar with AWS SageMaker, which offers Jupyter Notebook, allowing them to write and run code in one place. Starting week 2, we explored into the foundations of GPU computing, where students learned the anatomy of Compute Unified Device Architecture (CUDA) threads and blocks and performed hands-on experiments with Matrix operations using CuPy. As we progressed, students focused on data flow between the Central Processing Unit (CPU) (host) and GPU (device). They started profiling and used critical thinking and problem-solving skills to address memory bottlenecks, which impact performance. Before midterm, we introduced JIT libraries, as discussed, to exploit GPU parallelism for large datasets. After the midterm, students built Graph Convolutional Networks (GCNs) and trained them on a GPU; they explored various ways to use parallel programming concepts to aggregate node features on multiple GPUs for large-scale, real-world networks. Students were also required to build Retrieval Augmented Generation (RAG) systems and experiment with GPU-tuned retrievers and generators to optimize latency and throughput.
In this paper, we present the evaluation results of the course based on multiple instruments: students’ performance on assessments, their end-of-semester course evaluations, and two anonymous surveys (mid and post-course). Our takeaways from offering this course are: 1) AWS or any cloud resources are great and effective tools for infusing parallel programming with modern GPUs; 2) The best way to infuse HPC concepts to STEM students with minimal programming background is through hands-on learning with a mix of lecture and in-class labs;
3) Infusing HPC concepts into the curriculum to build AI agents has a positive impact on student learning, enabling them to apply these concepts to real-world data analysis tasks. To reinforce this learning, graded activities that require independent work should be designed as extensions of in-class labs, challenging students to apply their critical thinking and problem-solving skills. Therefore, we recommend the continued offering of this course and propose exploring its designation as a General Education course, with an introductory programming course as a prerequisite.

The remainder of the paper is organized as follows: Section II reviews related work, Section III outlines the course structure, Section IV presents evaluation results, and Section V concludes with future directions.
\section{Related Work}
The increasing pervasiveness and ubiquity of HPC systems, largely driven by the unprecedented growth and reliance on AI, requires the efficient execution of complex computational tasks. Consequently, GPUs have become indispensable components for optimizing the computing paradigm within AI applications, with major platforms such as Google Cloud, NVIDIA, and AWS SageMaker extensively utilizing GPU-accelerated and cloud-based computing resources to meet these demands. This technological evolution underscores the urgent need for academic curricula to incorporate HPC concepts, particularly parallel and GPU programming, to prepare students for current and emerging computational challenges.

Prasad et al.  proposed a practical framework as outlined in the NSF/IEEE TCPP Curriculum Initiative on Parallel and Distributed Computing, specifically the Core Topics for Undergraduates \cite{prasad2011nsf}. This initiative provides a nationally endorsed framework for embedding parallel and distributed computing concepts into undergraduate education. Chen et al. further support this initiative, highlighting the importance of computing competency, particularly in the context of developing practical skills for high-performance computing. Their support reflects a growing consensus that undergraduate computer science education must evolve to reflect the realities of modern computational environments\cite{chen2022producing}.

Arroyo echoes this need by promoting the integration of HPC and distributed programming across core computer science curricula. He proposes a progressive pedagogical framework that bridges sequential and parallel paradigms using a combination of traditional and novel teaching tools to enhance student preparedness \cite{arroyo2013teaching}. In alignment with Arroyo’s perspective, NVIDIA offers hands-on GPU-accelerated training programs in AI, HPC, and workflow automation that include teaching kits and hybrid learning formats tailored for university integration, supporting the importance of aligning academic instruction with evolving industry practices \cite{nvidia_training}.

Further advancing educational strategies, Qasem et al. present innovative approaches to parallel and distributed computing education that resonate with the need for practical GPU programming skills. Their work highlights GPU programming courses modeled on frameworks such as Carpentries, alongside specialized bootcamps focused on AI workloads in HPC systems, which collectively support hands-on learning and interdisciplinary student engagement \cite{qasem2022lightning}. In a related context, Adams et al. analyze the rapid adoption of cloud computing services and their implications for higher education curricula. They argue that the growing demand for practical, hands-on training in cloud computing, programming, and data science requires structured teaching materials, canonical learning objectives, and modular course design. Their approach ensures that cloud computing content can be seamlessly integrated into a variety of computing courses, facilitating student exposure to this rapidly evolving field \cite{adams2020cloud}.

Neelima et al. provide a case study of HPC integration into an undergraduate engineering curriculum over six years, documenting a progressive expansion from shared and distributed memory programming to accelerator and hybrid programming models. Their findings reveal that deliberate pedagogical strategies and institutional support can overcome initial challenges, ultimately yielding measurable benefits to both students and stakeholders \cite{neelima2015introducing}. Similarly, Qasem et al. advocate for a module-driven approach to embedding heterogeneous computing topics across the computer science curriculum. This strategy addresses curricular constraints by dispersing focused modules within existing courses, a method that demonstrated increased student engagement and foundational understanding of heterogeneous architectures pertinent to HPC, mobile processing, and Internet of Things (IoT) domains \cite{qasem2021module}.

Bridging theory and practice, Xu’s senior elective course effectively teaches OpenMP (Open Multi-Processing), Message Passing Interface (MPI), and CUDA using Google Colab and hands-on Raspberry Pi cluster construction to teach heterogeneous and parallel computing. Xu’s evaluation reports that this hybrid model significantly enhances students’ comprehension of data organization and parallel programming concepts, highlighting the value of experiential learning in HPC education \cite{xu2023teaching}.

In parallel, the increasing prominence of AI in diverse sectors has prompted curricular innovations beyond traditional computer science programs. Xu and Babaian address the challenges of delivering AI education to interdisciplinary and non-technical audiences by proposing a graduate-level curriculum that balances foundational AI knowledge with cutting-edge developments. Their study provides actionable guidelines for curriculum design, emphasizing the need for pedagogical approaches that engage students with varying backgrounds \cite{xu2021artificial}. 


Taken together, these works collectively argue for a curriculum that incorporates GPU programming and HPC as part of AI education, leveraging cloud-based platforms such as AWS SageMaker. The integration of GPU-accelerated cloud resources in academic training provides students with relevant, hands-on experience essential for navigating AI workflows in contemporary computational environments. Such curricular developments are vital to align educational outcomes with the technological demands of industry and research, ensuring that graduates are equipped to contribute effectively to AI-driven innovation.
\section{Course Structure}
\begin{figure}[htbp]
    \centering
    \includegraphics[width=0.75\columnwidth]{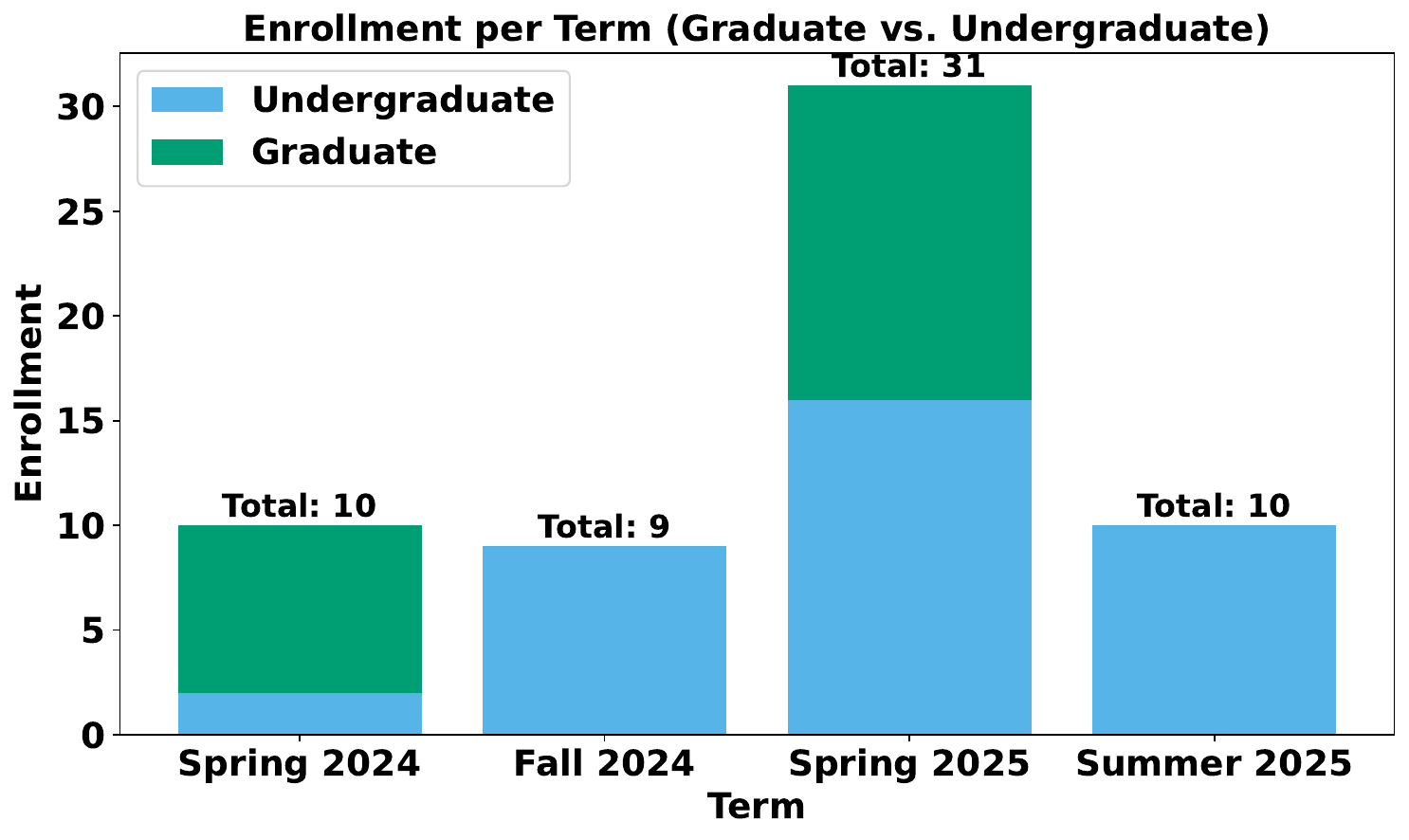}
    \caption{Enrollment per Term (Graduate vs. Undergraduate)}
    \label{fig:enrollment}
\end{figure}

This course plunges students into the fundamental concepts of accelerators and empowers them to leverage these powerful tools for creating sophisticated AI workflows and automated AI agents. Crucially, it integrates modules that introduce core HPC concepts, laying a robust foundation for modern AI development. The course is flexibly offered across the fall, spring, and summer semesters, contingent on enrollment. Each weekly 120-minute session is dynamically split: the initial half is dedicated to module lectures, while the subsequent half transforms into an immersive, hands-on lab. For spring and summer iterations, the course adopts a hybrid model, featuring optional in-person lab sessions to accommodate diverse student needs. This interdisciplinary offering caters to undergraduate seniors from Data Analytics, Information Systems, and Public Health Informatics programs, as well as graduate students in Information Systems. Designed as a comprehensive 16-week journey during the standard academic year, it condenses into an intensive four-week program during the summer. The detailed evaluation components are discussed in section IV. Figure \ref{fig:enrollment} showcases student enrollment across Fall 2024, Spring, and Summer 2025, the Spring 2025 iteration notably saw fifteen graduate students enroll. For the scope of this paper, we will exclusively analyze course evaluation and assessment results from the Fall 2024 and Spring 2025 sessions, as the Summer 2025 session is currently ongoing.
\vspace{-0.5em}
\subsection{Infrastructure Setup}
One of the cornerstones of our pedagogical approach was providing students with unparalleled hands-on access to AWS cloud resources. Each student was assigned a dedicated Identity and Access Management (IAM) role, empowering them to independently launch instances, configure NVIDIA GPUs, and provision the necessary computational power for their labs and assignments. We further augmented this practical experience by leveraging AWS Educate resources, training students not only on core AWS services but also on the intricacies of AWS SageMaker and spinning up new instances.
To ensure responsible resource utilization, each student's usage was capped for all assessments, complemented by automated scripts designed to terminate idle resources. Our current budget allows us to provision ample AWS resources until July 2026, ensuring sustained access. For efficient management and monitoring of unused instances, all GPU instances are provisioned within the US East (N. Virginia) region. Students were provided with a bootstrap script that simplified resource configuration using their AWS credentials for each assessment.
\subsubsection{AWS Cost}
The cost efficiency of this approach is significant: the average on-demand cost for a single-GPU instance was approximately \$1.262 per student per hour, while multi-GPU instances (up to 3) averaged about \$2.314 per student per hour. Over the entire semester, students typically spent around 40-45 hours utilizing AWS resources (including GPUs, AWS SageMaker, and VPC) for their assessments, translating to an average cost of roughly \$50-60 per student for the entire semester. Appendix A provides an overview of average AWS GPU usage and cost during Fall 2024 and Spring 2025.
For certain assessments, we strategically utilized AWS Educate resources, which are provided free of charge by AWS, further enhancing cost-effectiveness and accessibility. Furthermore, we allowed students to request additional resources, capped at \$100 per student for the semester; remarkably, no one found it necessary to request additional funds, highlighting the sufficiency of the allocated resources. While we also explored Google Colab, a platform frequently discussed in academic literature \cite{xu2023teaching}, its limitation to a single GPU per session made AWS a superior choice for the diverse and demanding multi-GPU AI workloads central to our curriculum.

\subsection{Modules}
Our primary aim is to immerse students in GPU programming concepts (GPC), enabling them to confidently navigate and utilize various powerful libraries such as RAPIDS for GPU-accelerated data science and Dask for scalable computing. Students are expected to gain a profound understanding of how GPU architecture fundamentally differs from CPU architecture, exploring into the core components of GPUs like Kernels, blocks, grids, and threads. 
Each course module was developed using a variety of resources, including academic peer-reviewed papers and official documentation from libraries such as Nvidia \cite{nvidia_training}, Dask \cite{dask_tutorial}, and PyTorch \cite{pytorch_profiler_tutorial}. The inclusion of academic papers as course materials was inspired by curriculum development articles on High-Performance Computing \cite{neuwirth2023bridging, kremer2023challenges}.
To support student learning, we provided access to lecture slides and supplementary reference materials for each module. The textbook \cite{dias2019hands} was used to introduce students to the fundamental concepts of GPU programming and architecture.

Crucially, all labs, assignments, and exams require students to employ a Python-based interface for GPU programming, rather than native CUDA (C/C++).
Our objective is to familiarize them with the architecture and empower them to create sophisticated AI workflows that are heavily supported by the rich ecosystem of Python libraries. While we acknowledge the inherent performance compromises of not using native CUDA for this course, our pedagogical focus remains on accessibility and practical application for this specific student demographic.

Table \ref{tab:modules_slo} outlines the modules covered during the 16-week course, alongside their corresponding student learning outcomes, and the deliverables for that week, which includes a weekly quiz for each module except week 7, and the final exam. 

Students engaged in the development of several distributed GPU programs, prominently featuring the partitioning of large-scale, real-world networks such as PubMed \cite{sen2008collective} and Reddit \cite{hamilton2017inductive}. This foundational step enabled the subsequent execution of Graph Convolutional Network (GCN) \cite{kipf2017semi} training for node label prediction across multiple GPUs. 

\algtext*{EndWhile}%
\algtext*{EndFor}%
\algtext*{EndIf}%
\algtext*{EndProcedure}%

\begin{algorithm}
\caption{Distributed GCN Training Using METIS Partitioning and Dask}
\label{algorithm:1}
\footnotesize
\begin{algorithmic}[1] 

\Require Undirected graph $\mathcal{G} = (V, E)$
\Require Node features matrix $X \in \mathbb{R}^{|V| \times d}$
\Require Label vector $Y \in \mathbb{R}^{|V|}$
\Require GCN model $M$ with parameters $\theta$
\Require Loss function $\mathcal{L}$ and optimizer $\mathcal{O}$
\Require Number of partitions $k$, training epochs $N_{epochs}$

\Ensure Trained model parameters $\theta$

\Procedure{TrainDistributedGCN}{$\mathcal{G}, X, Y, M, \mathcal{L}, \mathcal{O}, k, N_{epochs}$}
    \State Load $\mathcal{G}$, $X$, and $Y$; compute normalized adjacency matrix $\tilde{A}$
    \State Partition $\mathcal{G}$ into $\{\mathcal{G}_1, \dots, \mathcal{G}_k\}$ using METIS
    \State Initialize Dask cluster; assign each worker to a GPU

    \For{$i = 1$ to $k$}
        \State Distribute $\mathcal{G}_i$, $X_i$, $Y_i$ to worker $i$
    \EndFor

    \State Initialize global model $M_{global}$ with parameters $\theta$
    \State Broadcast $\theta$ to all workers

    \For{$epoch = 1$ to $N_{epochs}$}
        \For{$i = 1$ to $k$}
            \State On worker $i$: compute local loss and gradients
        \EndFor
        \State Aggregate gradients from all workers
        \State Update global model parameters $\theta$ using $\mathcal{O}$
        \State Report epoch loss
    \EndFor

    \State \Return Trained parameters $\theta$
\EndProcedure

\end{algorithmic}
\end{algorithm}

Algorithm~\ref{algorithm:1} illustrates a classic node label problem. Here, a GCN model processes node features, $X \in \mathbb{R}^{|V| \times d}$, and the graph's structural information, typically represented by its adjacency matrix, derived from the graph $\mathcal{G} = (V, E)$. The model's objective is to predict labels for nodes, $Y \in \mathbb{R}^{|V|}$. A key strategy for infusing parallel processing into this workflow involves distributing the graph across multiple GPUs, allowing each device to train on a distinct subgraph. As delineated in Algorithm~\ref{algorithm:1}, line 3, the graph is initially partitioned using the highly efficient METIS algorithm \cite{karypis1998metis}. Subsequently, lines 5 and 6 detail the distribution of these subgraphs across the available GPUs, facilitating parallel training (lines 9-10). We specifically leveraged the Dask framework \cite{rocklin2015dask} for orchestrating this distributed training process. Within this parallel paradigm (lines 10-12), each GPU worker independently processes its assigned partition, computes the local loss by comparing predicted node labels with ground truth, and calculates gradients. These local gradients are then efficiently combined, and a global optimizer updates the model's parameters in a synchronized fashion. 

While students were encouraged to explore and devise their own orchestration methods for graph splitting and parallel training, our work illuminated several interesting challenges inherent in distributed graph processing. Furthermore, to foster a deeper understanding of partitioning effects, students were encouraged to experiment with random graph partitioning as an alternative to METIS \cite{karypis1998metis} and to thoroughly analyze the resulting GPU utilization patterns.

Parallelizing large-scale, often dense graphs presents significant difficulties. In most cases, we observed that simply splitting the graph and distributing the training yielded minimal performance improvement. However, a notable outcome was the enhanced prediction accuracy scores after splitting and training, particularly when compared to sequential approaches.
\begin{table*}[htbp]     
    \centering 
    \caption{\label{tab:modules_slo}Course Modules, Student Learning Outcomes (SLOs), and Deliverables}     
    \begin{tabular}{|p{0.25\textwidth}|p{0.45\textwidth}|p{0.25\textwidth}|}         
        \hline
        \textbf{Week / Topic} & \textbf{Student Learning Outcome (SLO)} & \textbf{Deliverable} \\         
        \hline
        \textbf{Week 1: AWS GPU Setup + Course Introduction} & \textbf{Apply}: Set up AWS EC2 GPU instances and configure Python environments & Lab 1: AWS GPU instance setup with Jupyter and SSH access \\         
        \hline
        \textbf{Week 2: CUDA Fundamentals \& GPU Parallelism} & \textbf{Understand/Apply}: Explain GPU architecture, grasp CUDA programming basics, and implement parallel execution & Lab 2: CuPy vector/matrix operations \& parallel processing \\         
        \hline
        \textbf{Week 3: Memory Management \& GPU Optimization} & \textbf{Analyze/Optimize}: Manage and optimize memory transfers between host and GPU & Lab 3: Matrix multiplication with memory profiling using Numba \\         
        \hline
        & & Assignment 1: GPU Matrix Multiplication and Profiling (Due Week 5) \\         
        \hline
        \textbf{Week 4: GPU Profiling Tools \& Bottleneck Analysis} & \textbf{Analyze/Evaluate}: Apply Nsight Systems, PyTorch profiler, and cProfile for comprehensive GPU workload analysis & Lab 4: Profiling GPU RL loop with Nsight and PyTorch profiler \\         
        \hline
        & & Assignment 2: Distributed GPU Data Processing (Due Week 7) \\         
        \hline
        \textbf{Week 5: Custom CUDA Kernels with Python} & \textbf{Create/Integrate}: Write, compile, and seamlessly integrate custom CUDA kernels in Python workflows & Lab 5: Custom CUDA kernel with Numba + profiling \\         
        \hline
        \textbf{Week 6: RAPIDS + Dask for Scalable Data Pipelines} & \textbf{Apply/Create}: Process large datasets efficiently using RAPIDS cuDF and Dask for distributed GPU workflows & Lab 6: Parallel data processing using Dask with RAPIDS cuDF \\         
        \hline
        \textbf{Week 7: Midterm Exam / Assessment} & No SLO (Assessment Week) & Midterm Exam; Assignment 2 Due \\         
        \hline
        \textbf{Week 8: Deep Learning on GPUs (PyTorch Focus)} & \textbf{Apply/Optimize}: Train and optimize neural networks using GPU acceleration, specifically focusing on GCNs & Lab 7: CNN model training on GPU using PyTorch \\         
        \hline
        \textbf{Week 9: Reinforcement Learning on GPUs} & \textbf{Develop/Implement}: Develop reinforcement learning agents accelerated by GPUs & Lab 8: DQN agent training using CUDA-enabled PyTorch \\         
        \hline
        \textbf{Week 10: Multi-GPU Training \& Parallel Strategies} & \textbf{Apply/Scale}: Scale models efficiently using multi-GPU setups with Distributed Data Parallel (DDP) & Lab 9: PyTorch DDP implementation across 2 GPUs \\         
        \hline
        \textbf{Week 11: AI Agent Foundations \& GPU Benefits} & \textbf{Understand/Describe}: Describe AI agents and explain the GPU's critical role in training acceleration & Lab 10: Simple reinforcement agent using CuPy/Numba \\         
        \hline
        & & Assignment 3: Multi-GPU AI Agent (Due Week 13) \\         
        \hline
        \textbf{Week 12: Retrieval-Augmented Generation (RAG) Basics} & \textbf{Understand/Describe}: Describe RAG architectures, combining retrieval and generation modules effectively & Lab 11: Basic RAG pipeline using FAISS for retrieval \\         
        \hline
        \textbf{Week 13: GPU-Optimized RAG Development} & \textbf{Construct/Optimize}: Construct and optimize RAG models using GPU-accelerated retrievers and generators & Lab 12: Build GPU-enabled RAG with retriever + small LLM \\         
        \hline
        \textbf{Week 14: RAG Pipeline Optimization \& Inference} & \textbf{Optimize/Deploy}: Optimize end-to-end RAG pipelines for efficient real-time GPU inference & Lab 13: Deploy real-time RAG inference pipeline \\         
        \hline
        & & Assignment 4: End-to-End RAG System (Due Week 16) \\         
        \hline
        \textbf{Week 15: Project Development \& Support} & \textbf{Apply/Create}: Apply GPU acceleration, AI agent techniques, and RAG models in capstone projects & Lab 14: Build your own Lab (Extra Credit); Academic paper review (Extra Credit) \\         
        \hline
        \textbf{Week 16: Final Project Presentations \& Exam} & \textbf{Showcase/Demonstrate}: Showcase final projects demonstrating GPU-accelerated AI/RAG pipelines & Final Project Presentation \\         
        \hline
    \end{tabular} 
\end{table*}

\section{Evaluation Instruments and Results}
We utilize a multifaceted assessment approach, incorporating students' grades, their end-of-semester course evaluations, the comprehensive course project, and two anonymous surveys (mid- and post-course) to thoroughly gauge the effectiveness of our course modules and pedagogy.
\subsection{Student Performance Assessment}
Our rigorous assessment framework includes twelve to fourteen dynamic in-class labs (flexibly adjusted to class pace), four challenging assignments, and a collaborative group project (capped at two members). The project that constitutes 15\% of final grade. We also evaluate attendance and class participation.
Class participation is assessed through the submission of scribed notes for each lecture, along with a thoughtful question related to the module covered. Both the notes and the question are due immediately after class. This participation component is complemented by two comprehensive exams, a midterm and a final.

While all grading activities contribute to the final grade, their weighting is intentionally uneven. For labs and assignments, students will have direct access to both the instructor and the Teaching Assistant (TA), fostering a supportive learning environment where assistance is readily available. These highly interactive activities collectively constitute half of final grade, emphasizing hands-on application. The remaining half comes from independent endeavors, including the assignments and closed-book exams, designed to assess individual mastery, along with the group project.

\begin{figure}[htbp]
    \centering
   \includegraphics[width=\linewidth, keepaspectratio]{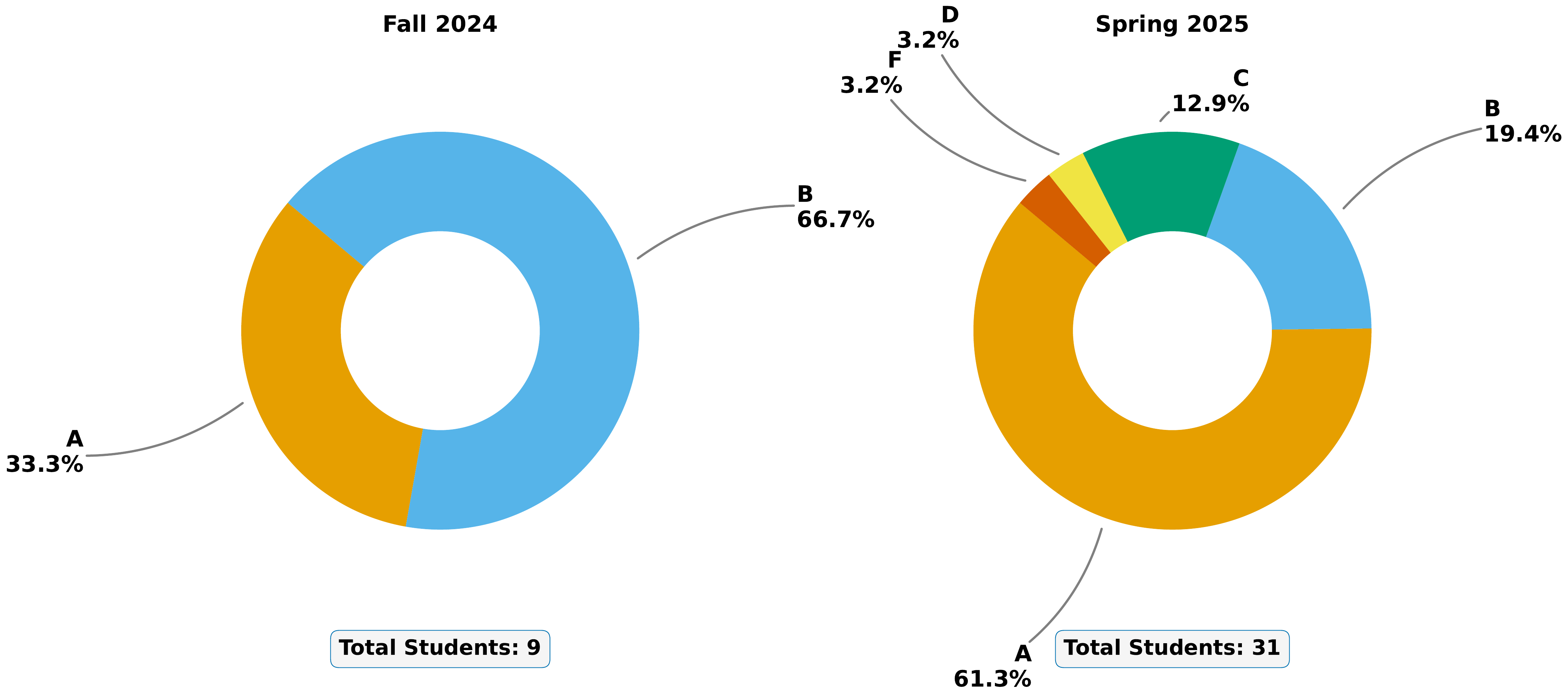}
    \caption{Grade Distribution for Fall 2024 and Spring 2025 Course Offerings}
    \label{fig:grade_distribution}
\end{figure}

Appendix B lists the extra credit opportunities available to students.
Figure \ref{fig:grade_distribution} visually depicts the grade distribution across Fall 2024 and Spring 2025. In Fall 2024, the majority of students achieved a 'B' grade, with many encountering  difficulties with post-midterm modules. Despite receiving assistance from the TA and instructor, a number of students were only able to submit partial assignments, suggesting challenges in independent completion. In stark contrast, Spring 2025 witnessed a significant uplift, with over 60\% of students securing an 'A' grade and demonstrating enhanced independence in completing assignments. This marked improvement is directly correlated with proactive revisions to our lab instructions, meticulously designed to better prepare students for subsequent assignments. We further observed that a substantial majority of students proactively leveraged office hours throughout the semester, not only to clarify complex assignments but also to request crucial code reviews prior to submission. Conversely, students earning a 'B' or lower typically correlated with missed submissions or late lab/assignment turn-ins.
The exam average remained remarkably consistent across both semesters, hovering between 75-80\%. This reflects a solid foundational understanding among the students. (See Appendix C for further data analysis and findings).
\subsection{End-of-Semester Student Evaluations}
\begin{figure*}[!htbp]
    \centering
   \includegraphics[width=\linewidth, keepaspectratio]{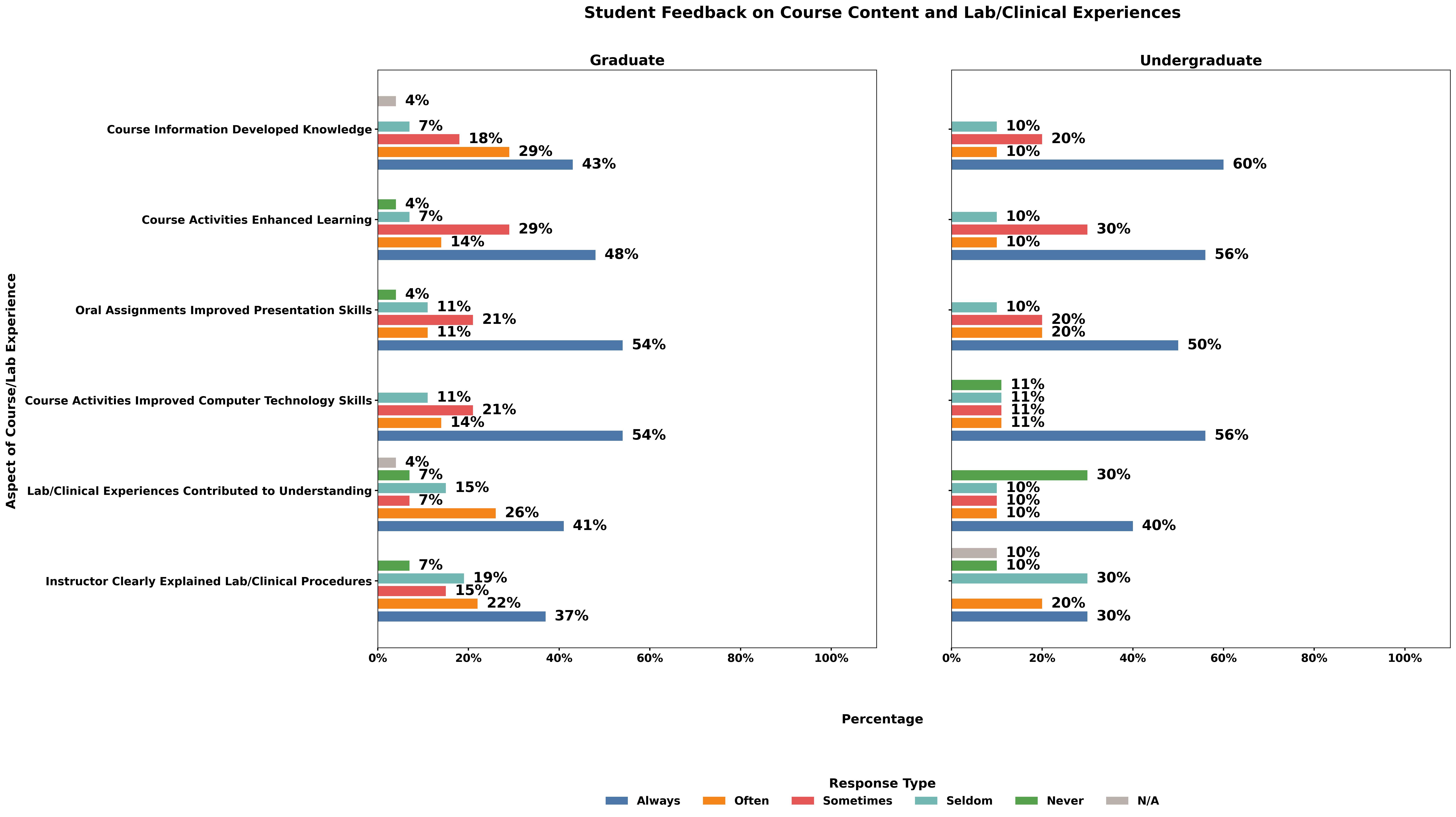}
    \caption{Student Feedback on Course Content and Lab/Clinical Experiences}
    \label{fig:student_feedback}
\end{figure*}

A robust 85\% of students completed the anonymous online evaluation form, providing valuable feedback on the course using questions presented in Table \ref{tab:evaluation_questions}. All questions, which are standard for all courses and designed by the university, utilized a five-point Likert scale with response options including "Always," "Often," "Sometimes," "Seldom," "Never," and "N/A."
\begin{table}[htbp]
\centering
\caption{\label{tab:evaluation_questions}End-of-Semester Course Assessment Questions}
\begin{tabular}{p{0.95\columnwidth}} 
\toprule
\textbf{Evaluation Questions} \\
\midrule
The course information further developed my knowledge in this area. \\
The course activities enhanced my learning of the course content. \\
The oral assignments improved my presentation skills. \\
The course activities improved my computer technology skills. \\
Lab or clinical experiences contributed to my understanding of the course theories and concepts. \\
The instructor clearly explained laboratory or clinical experiments or procedures. \\
\bottomrule
\end{tabular}
\end{table}
Figure \ref{fig:student_feedback} provides an overview of the student feedback regarding course content, with the X-axis representing the percentage of responses for each type and the Y-axis representing the questions from Table \ref{tab:evaluation_questions}. 
Figure \ref{fig:student_feedback} suggests that both undergraduate and graduate courses are largely successful in delivering positive learning experiences. However, there appear to be subtle differences in the perceived strengths, with undergraduates valuing core course content and graduates finding more significant gains in specific skill development. Further investigation into the methodologies and learning objectives of these course levels could provide deeper insights into these observed differences. The "Seldom," "Never," and "N/A" categories consistently represent a small minority of responses, indicating that strongly negative or irrelevant feedback is not prevalent. For both groups, "Lab/Clinical Experiences Contributed to Understanding" and "Instructor Clearly Explained Lab/Clinical Procedures" tend to have lower "Always" percentages compared to the course content aspects, suggesting these might be areas for instructors or course designers to explore for enhancement.
We plan to experiment in Fall 2025 by revising our labs and offering a more clear explanation about labs and assignments during class sessions and to explicitly highlight how assignments serve as extensions of the lab experiences. Although the survey question wording is standardized across disciplines, in this course “lab/clinical procedures” specifically refers to programming and computational tasks rather than healthcare contexts. (See Appendix D for further data analysis on student course satisfaction.)
\vspace{-0.5em}


\subsection{Anonymous Surveys}
During the Fall 2024 and Spring 2025 semesters, students were invited to provide anonymous feedback. This feedback was crucial for the department to evaluate existing course offerings and strategically plan for future similar courses, particularly concerning GPU programming content and pedagogical approaches. Feedback was systematically collected via Google Forms at two key points: during the sixth week (prior to the midterm exam) and the twelfth week (before students began their group projects). The primary objective of these two surveys was to ascertain the extent to which students had grasped the course material. Each survey question utilized a five-point Likert scale. 
Participation in the surveys was robust, with most students completing them. The mid-course survey featured three questions assessing their confidence in using Numba, building and configuring GPUs on AWS, and utilizing HPC tools for profiling. The final survey repeated these questions and additionally included a question on their confidence level in using multiple GPUs for parallel algorithm implementation. 

\begin{figure}[htbp]
    \centering
    \begin{subfigure}{0.4\textwidth}
\includegraphics[width=\linewidth, keepaspectratio]{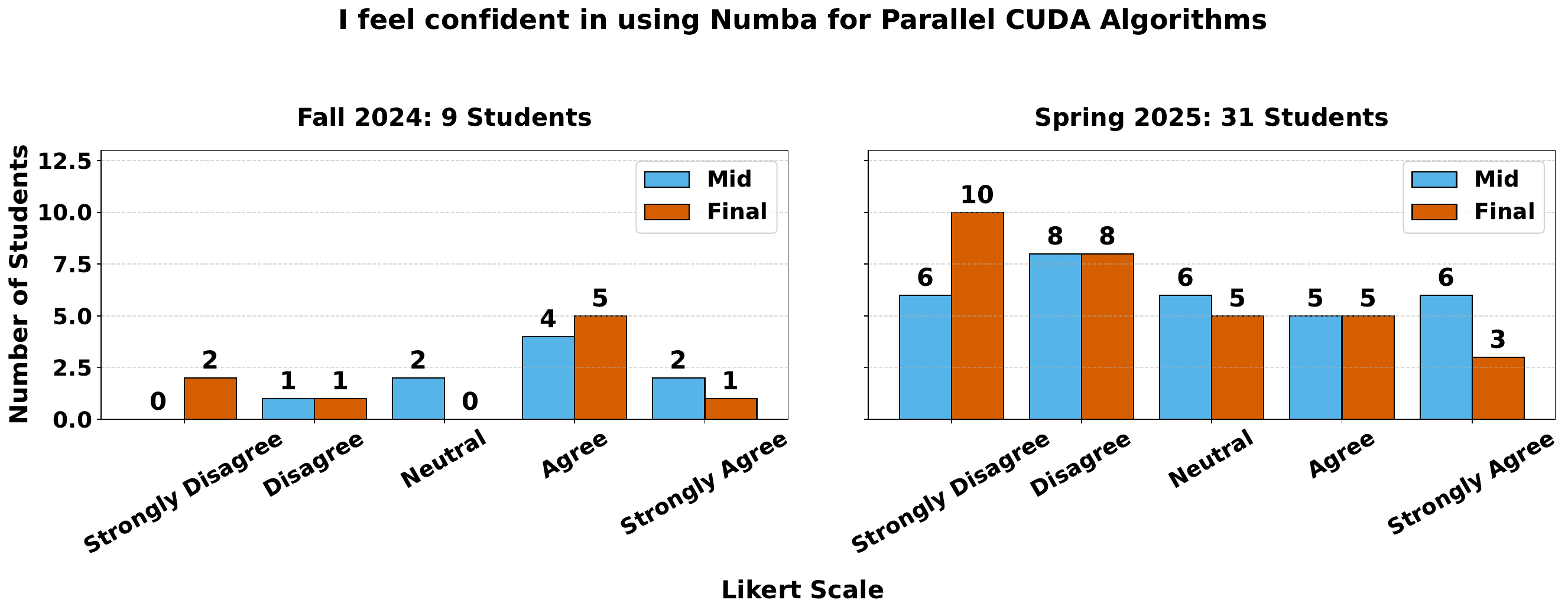}
        \caption{Numba-CUDA Parallel Programming Ability}
        \label{fig:numba_cuda_self_assessment}
    \end{subfigure}
    \hfill
    \begin{subfigure}{0.4\textwidth}
       \includegraphics[width=\linewidth, keepaspectratio]{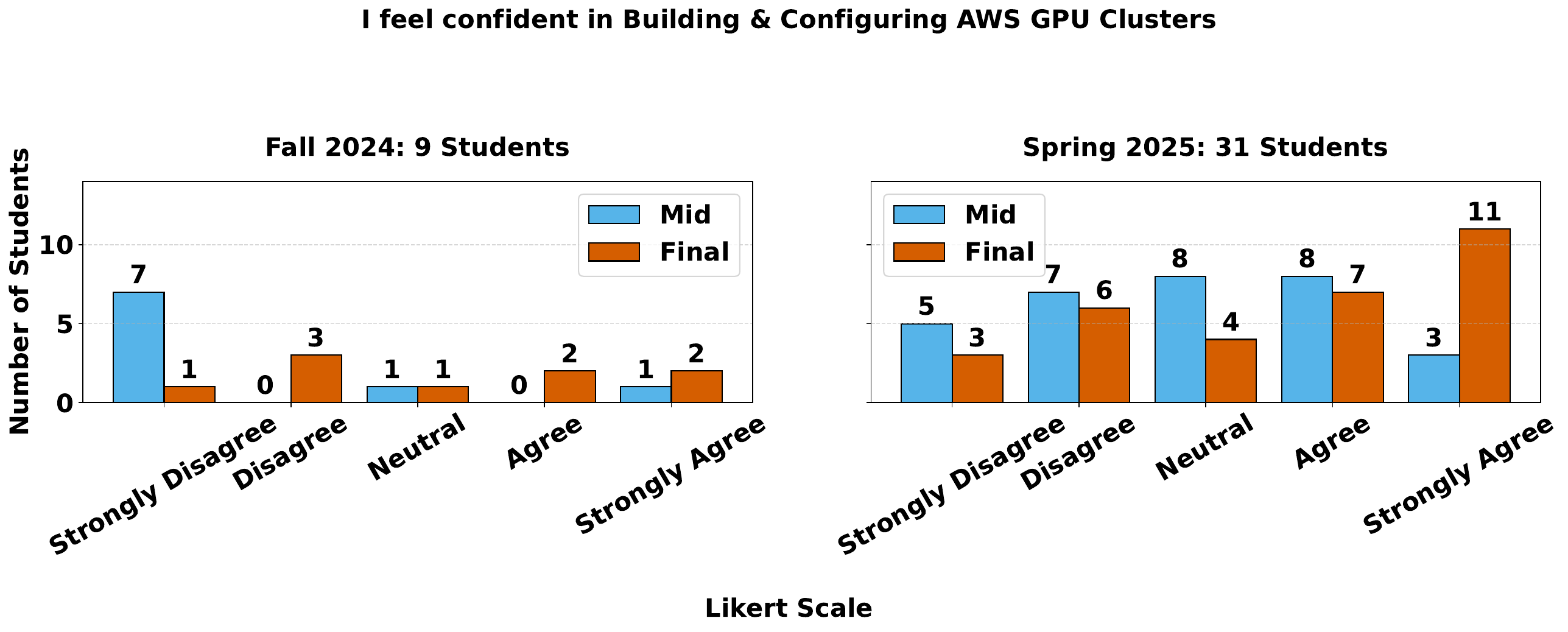}
        \caption{Confidence in Using AWS GPU Cluster}
        \label{fig:aws_gpu_cluster_confidence}
    \end{subfigure}

    \vspace{0.8em} 

    \begin{subfigure}{0.4\textwidth}
        \includegraphics[width=\linewidth,keepaspectratio]{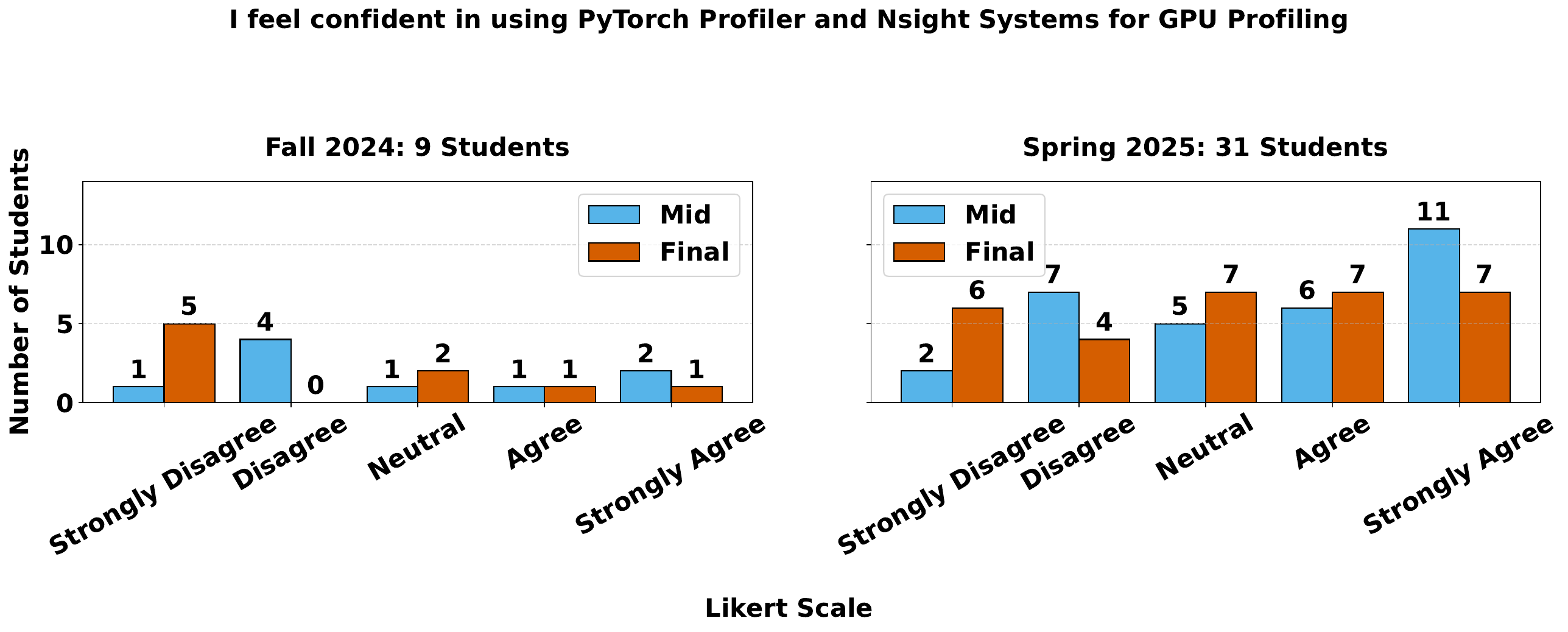}
        \caption{Python/PyTorch and Nsight GPU Profiling Confidence}
        \label{fig:python_pytorch_nsight_self_assessment}
    \end{subfigure}
    \hfill
    \begin{subfigure}{0.4\textwidth}
        \includegraphics[width=\linewidth, keepaspectratio]{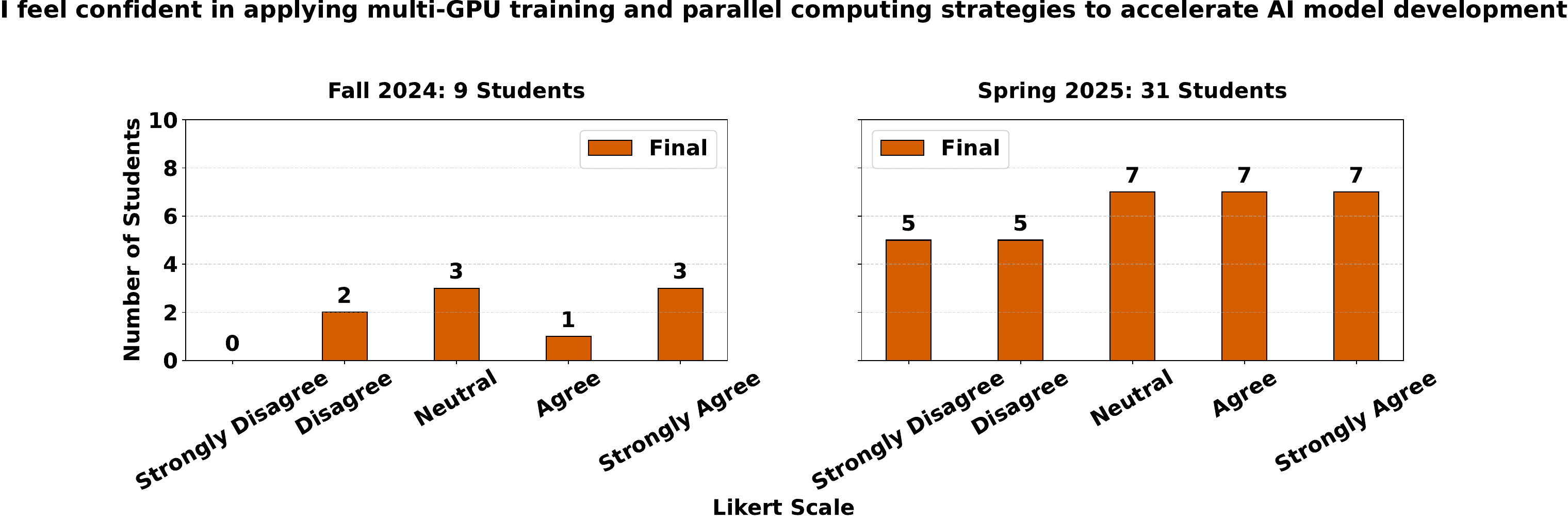}
        \caption{Multi-GPU Programming Confidence}
        \label{fig:multi_gpu_confidence}
    \end{subfigure}
    \caption{Student survey results (Fall 2024 / Spring 2025)}
    \label{fig:all_confidence_results}
\end{figure}

Figure \ref{fig:numba_cuda_self_assessment} compares student self-assessment of their ability to use Numba to implement a parallel algorithm using CUDA across two semesters. In Fall 2024, responses were relatively evenly distributed: two students strongly disagreed, two disagreed, one was neutral, two agreed, and two strongly agreed.
Conversely, Spring 2025 shows a greater proportion of students in the 'Neutral' to 'Strongly Agree' categories. Specifically, nine students were neutral, seven agreed, and five strongly agreed, making 'Neutral' the largest single response group. This suggests that while many students felt capable, a significant portion remained in the middle, indicating they felt neither fully proficient nor entirely lacking in this specific skill.

Figure \ref{fig:aws_gpu_cluster_confidence} illustrates student confidence level in building and configuring GPU clusters. In Fall 2024, students initially expressed weak confidence during the midterm. We believe this stemmed from initial challenges in configuring GPUs and ensuring instances were correctly connected within the same Virtual Private Cloud (VPC) with appropriate subnet addresses. By the final survey, confidence improved; we attribute this to the automation of instance setup and networking, although some students still faced challenges in updating and executing these scripts correctly.

In Spring 2025, midterm confidence was mixed, with approximately twelve students expressing disagreement, eight remaining neutral, and eleven showing agreement. By the final survey, confidence had substantially improved, with students reporting strong confidence. We posit this increase in confidence resulted from the reinforcement of cluster creation tasks throughout all in-class activities. 

Figure \ref{fig:python_pytorch_nsight_self_assessment} illustrates student confidence level in utilizing PyTorch Profiler \cite{pytorch_profiler_tutorial} and Nsight Systems\cite{nvidia_nsight_systems} for GPU profiling. In Fall 2024, students initially expressed strong confidence during the midterm assessment; however, a clear reduction in confidence was observed in the final survey regarding their proficiency with GPU profiling tools. We suggest this decline may be attributed to the increased independent application of these profilers for project work after the midterm. A similar trend was observed in Spring 2025; however, the magnitude of the confidence dip (from "agree" to "disagree") was less pronounced compared to Fall 2024. This attenuated decline in Spring 2025 might be a consequence of incorporating additional hands-on profiling activities during in-class sessions.
Figure \ref{fig:multi_gpu_confidence} presents final survey results on student confidence in applying multi-GPU training and parallel computing for AI models such as GCN. Fall 2024 responses, from a small group of students, were largely positive. Spring 2025 results, from a larger group, were more varied, with ten students expressing disagreement while most reported neutral or higher confidence. The moderate confidence may reflect the assignment’s difficulty, particularly issues installing PyTorch and configuring Dask (Algorithm \ref{algorithm:1}).

\section{Conclusions}
Despite the rapid advancements in AI and the growing accessibility of GPU technologies, integrating GPU programming and HPC concepts into the STEM curriculum remains a significant challenge. We contend that the most effective pedagogical approach to bridge this gap involves hands-on laboratory exercises and assessments that not only impart technical skills but also cultivate critical thinking and problem-solving abilities among STEM students.

We successfully introduced GPU programming. Our experience in delivering these special topics courses, as detailed in this paper, has been overwhelmingly positive. These offerings have provided invaluable insights into student engagement and learning outcomes.

Looking ahead, we plan to refine and expand our lab modules and assessments based on these experiences. We are actively exploring various ways to integrate specific course modules into other Data Analytics offerings. To better prepare students for these advanced topics, we are considering a revision of the prerequisite course to infuse foundational HPC concepts alongside traditional sequential programming.
Furthermore, we intend to explore alternative pedagogical strategies and continuously evolve the curriculum to align with the AI and HPC competencies currently in high demand in the job market. Our overarching goal is to offer this course more frequently, making it accessible to a broader and more diverse STEM student population.
\bibliographystyle{IEEEtran} 
\bibliography{main} 
\clearpage
\appendices
\section{AWS GPU Computational Hours}
\begin{figure}[htbp]
    \centering
    \includegraphics[width=0.8\columnwidth, ]{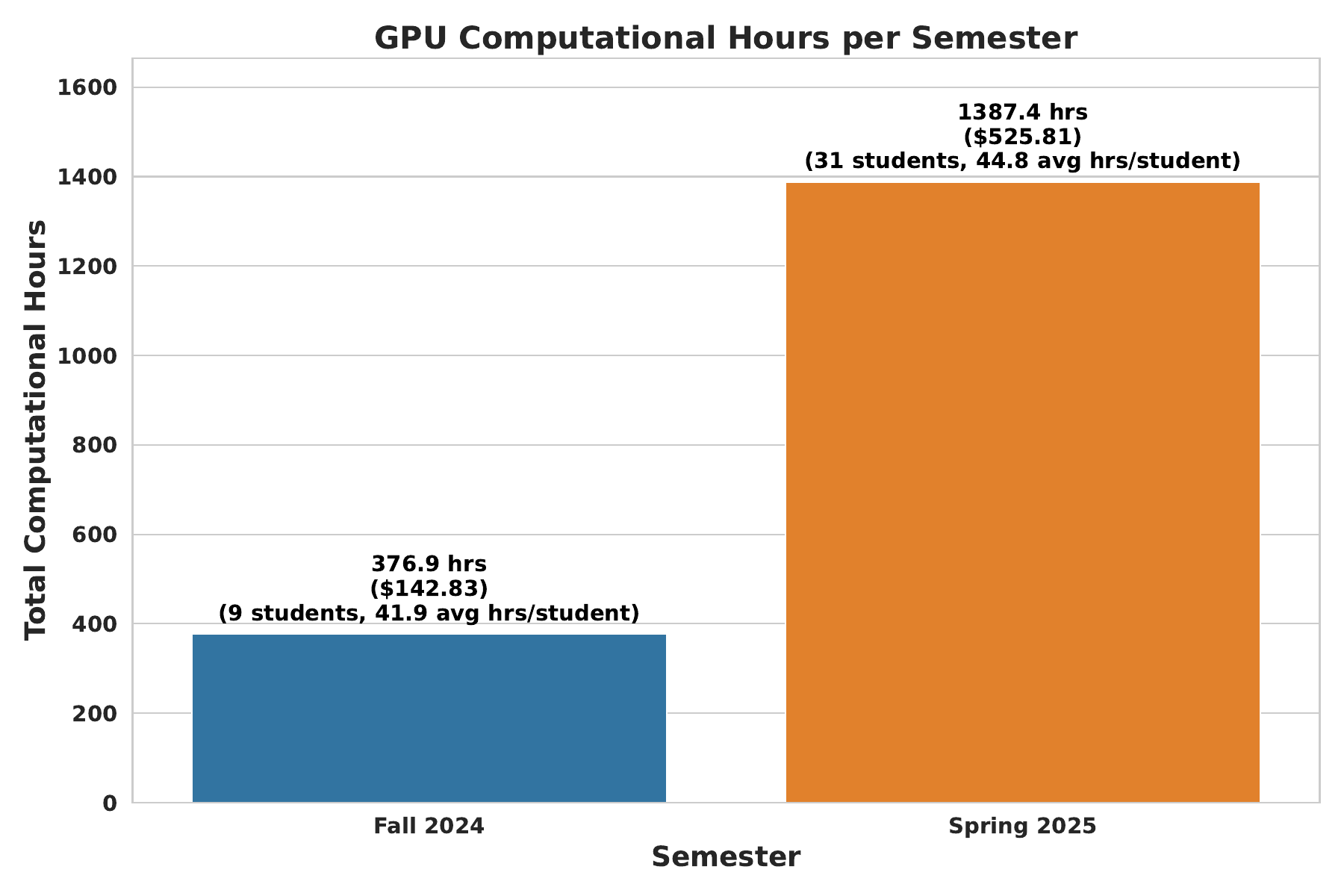}
    \caption{Average AWS GPU usage and cost for Fall 2024, and Spring 2025.}
    \label{fig::Awscomputation}
\end{figure}
\vspace{-0.1em}
Figure \ref{fig::Awscomputation} provides an estimated average computational hours of GPU instances for all labs and assignments in Fall 2024 and Spring 2025. The notable increase in average hours per student during Spring 2025 is due to the introduction of two additional labs. For group projects, the average usage of GPU resources was less than 2 hours in both semesters. We did not include the computational hours of GPU instances from AWS Educate because the instructor lacks access to resource usage insights for that platform.

\section{Extra Credit}

Two distinct extra credit opportunities were offered to students.

\textbf{1. Build Your Own Lab}

Students were encouraged to design their own lab based on the course modules. The only requirement was that the lab could not replicate an existing one. The primary goal of this initiative was to assess students' ability to apply core concepts and design a lab from scratch. A secondary objective was to identify potential labs that could be rigorously tested during the summer of 2025 and possibly integrated into future semesters. No students attempted this extra credit in Fall 2024. In Spring 2025, three students submitted the lab; however, none of the submissions fully met the student learning outcomes. We believe this may be due to students attempting the design and preparation during finals week, which prevented them from dedicating sufficient time to research and development.

\textbf{2. Academic Paper Review}

In Spring 2025, an extra credit opportunity was offered where students could select a peer-reviewed academic paper published between 2020 and 2025 based on the modules discussed. They were asked to provide a one-page summary, discuss its shortcomings, and propose an extension of the research. The goal was to help students understand the current research landscape in their chosen modules. The inclusion of this extra credit was inspired by an article\cite{neuwirth2023bridging}; where attempt was made to bridge the gap between education and research. Approximately 60\% of students completed this activity. While most provided excellent summaries, their explanations for expanding on the proposed research were often vague. Based on this outcome, we plan to explore integrating academic papers directly into course modules to better promote student research skills.

\section{Data Analysis and Findings}
We evaluated potential differences in academic performance between graduate and undergraduate students enrolled in a special topics course, specifically focused on GPU programming, offered in Fall 2024 and Spring 2025. To assess this, we tested the following hypotheses:

\begin{itemize}
    \item \textbf{Null Hypothesis ($H_0$)}: There is no difference in average performance between graduate and undergraduate students.
    \item \textbf{Alternative Hypothesis ($H_1$)}: There is a difference in average performance between graduate and undergraduate students.
\end{itemize}

As we did not specify a direction for the difference, a two-tailed test was employed to assess whether the observed performance differences reflected random variation or a systematic distinction between the two cohorts. Before conducting hypothesis testing, we evaluated whether the data satisfied the assumptions required for parametric tests—namely, normality and homogeneity of variance.

\begin{figure}[htbp]
    \centering \includegraphics[width=0.8\columnwidth]{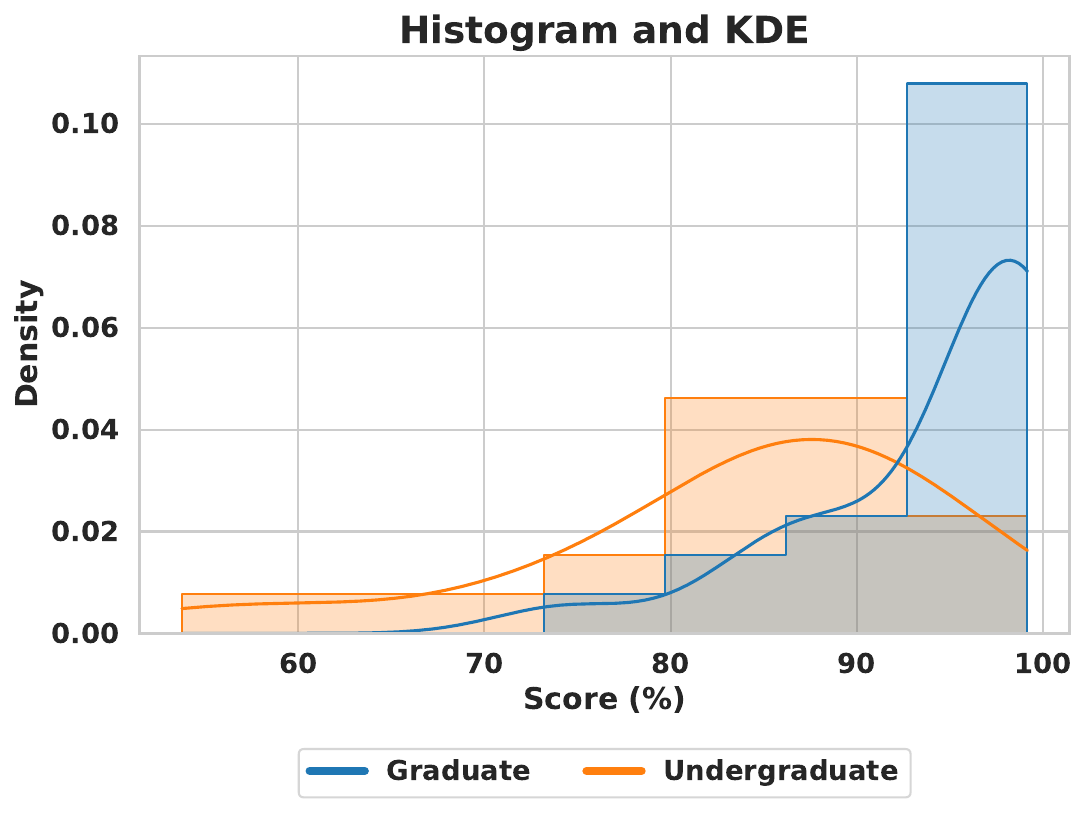}
    \caption{Histogram comparison of academic scores between graduate and undergraduate student groups.}
    \label{fig:histogramsplots}
\end{figure}

\begin{figure}[htbp]
    \centering
    \begin{minipage}[c]{0.48\columnwidth}
        \centering
        \includegraphics[width=\columnwidth]{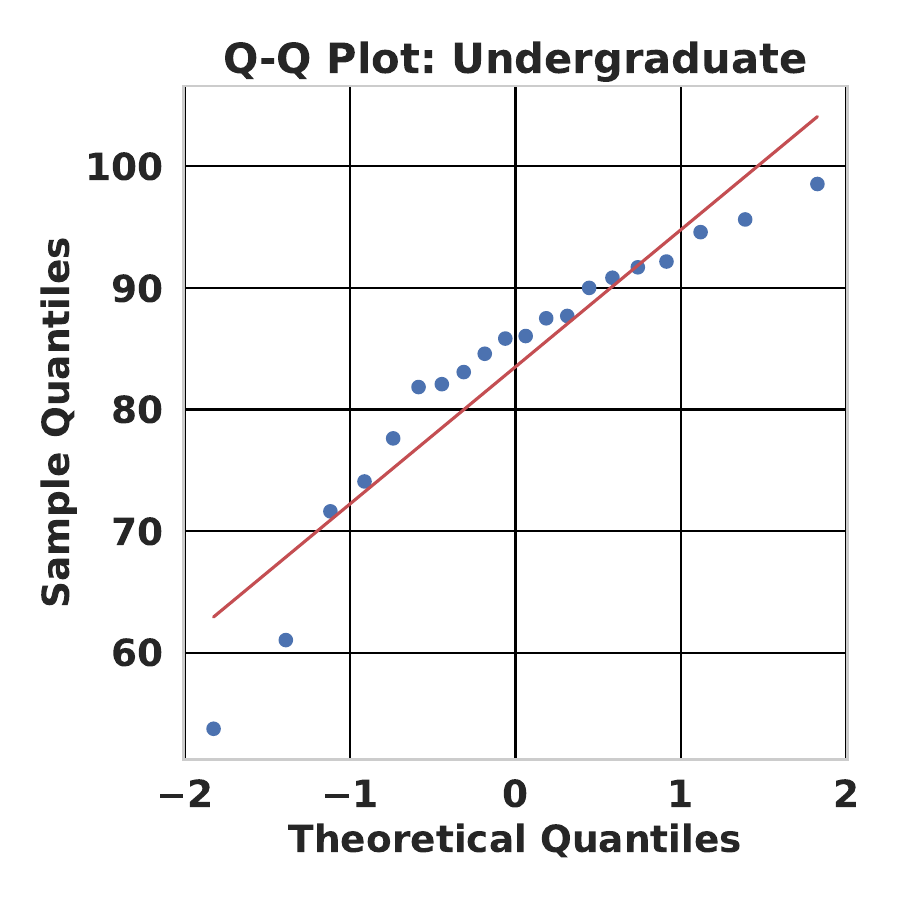}
        \caption{Q–Q plots of academic scores for undergraduate students.}
        \label{fig:qqundergraduateplots}
    \end{minipage}\hfill
    \begin{minipage}[c]{0.48\columnwidth}
        \centering
        \includegraphics[width=\columnwidth]{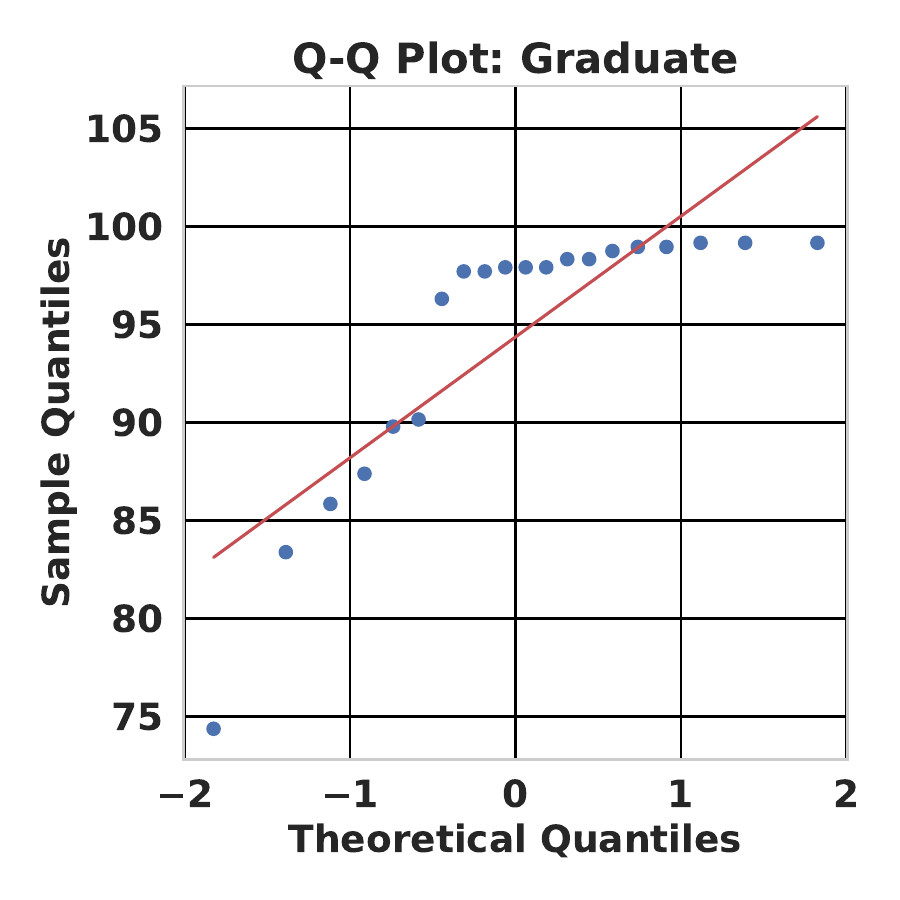}
        \caption{Q–Q plots of academic scores for graduate students.}
        \label{fig:qqgraduateplots}
    \end{minipage}
\end{figure}

Figure~\ref{fig:histogramsplots} displays histograms, while  Figure~\ref{fig:qqundergraduateplots} and Figure~\ref{fig:qqgraduateplots} present  Q--Q plots. These visualizations indicate clear departures from normality, particularly in the graduate group, whose scores were tightly clustered near the upper end of the distribution and exhibited noticeable skewness.

To statistically assess normality, we conducted Shapiro--Wilk tests. The graduate group showed a substantial deviation from normality ($W = 0.722$, $p < .001$), while the undergraduate group also deviated, though less severely ($W = 0.898$, $p = .037$). In contrast, Levene’s test for equality of variances revealed no significant difference between the two groups ($F = 2.437$, $p = .127$), suggesting that the assumption of homogeneity of variance was met. These results are summarized in Table~\ref{tab:normality}.

\begin{table}[ht]
\centering
\caption{Results of Assumption Tests for Normality and Homogeneity of Variance}
\label{tab:normality}
\begin{tabular}{lcc}
\hline
\textbf{Assumption Test} & \textbf{Statistic} & \textbf{p-value} \\
\hline
Shapiro--Wilk (Graduate) & 0.722 & $< .001$ \\
Shapiro--Wilk (Undergraduate) & 0.898 & .037 \\
Levene’s Test & 2.437 & .127 \\
\hline
\end{tabular}
\end{table}
\vspace{-0.2em}
In addition to assumption testing, descriptive statistics provided insight into the distribution and central tendency of scores in each group, as shown in Table~\ref{tab:descriptives}. Graduate students achieved remarkably higher mean scores and demonstrated a more compact distribution, reflected in lower variability compared to undergraduates.



\vspace{-0.5em}

\begin{table}[htbp]
\centering
\caption{Descriptive statistics for academic performance scores by group}
\label{tab:descriptives}
\resizebox{\columnwidth}{!}{%
\begin{tabular}{lcccccccc}
\hline
\textbf{Group} & \textbf{Mean} & \textbf{Std Dev} & \textbf{Min} & \textbf{Q1} & \textbf{Median} & \textbf{Q3} & \textbf{Max} & \textbf{Count} \\
\hline
Graduate & 94.36 & 6.91 & 74.38 & 90.06 & 97.92 & 98.80 & 99.17 & 20 \\
Undergraduate & 83.51 & 11.33 & 53.75 & 80.79 & 85.94 & 91.05 & 98.54 & 20 \\
\hline
\end{tabular}
}
\end{table}

\begin{figure}[htbp]
    \centering
    \includegraphics[width=1.0\columnwidth, ]{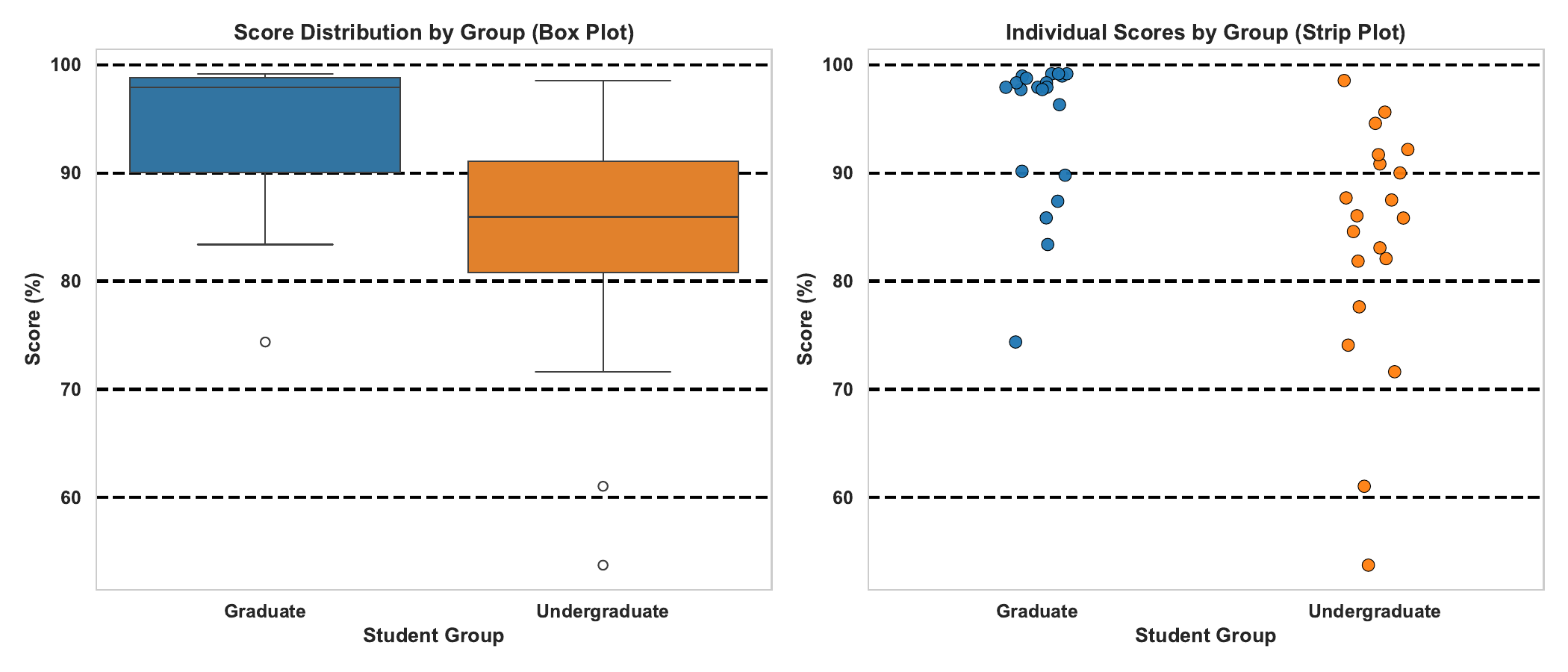}
    \caption{Boxplot and Stripplot of academic scores for graduate and undergraduate student groups.}
    \label{fig:stripplot_boxplot_side_by_side}
\end{figure}

Although the variances were homogeneous, the pronounced non-normality, especially in the graduate scores, made parametric tests (e.g., independent samples \textit{t}-test) inappropriate. Given the relatively small sample sizes ($n = 20$ per group) and non-normal distributions, we employed the Mann--Whitney \textit{U} test, a non-parametric alternative suitable for comparing central tendencies between two independent groups. The Mann--Whitney \textit{U} test revealed a statistically significant difference in performance between the two groups, with graduate students outperforming undergraduates ($U = 332.00$, $p = .0004$).
Figure~\ref{fig:stripplot_boxplot_side_by_side} which show a higher median and a more compact score distribution among graduate students compared to undergraduates.

In conclusion, the data provide strong evidence to reject the null hypothesis. Graduate students performed significantly better than undergraduate students in terms of their weighted total scores. By testing assumptions and applying a suitable non-parametric method, we ensured the validity of our findings and aligned our analytical approach with the data's distributional characteristics.

\section{Descriptive Statistics of Student Satisfaction}
We analyzed student satisfaction for the GPU programming course across Fall 2024 and Spring 2025 based on course evaluations collected anonymously ($n = 18$) and found overall positive results with some differences between the semesters. In Fall 2024, most students (87.5\%) reported ``Very High'' satisfaction, while one student (12.5\%) reported ``Very Low'' satisfaction. In Spring 2025, 60\% of students rated their satisfaction as ``Very High,'' and the remaining 40\% rated it as ``High,'' with no ``Very Low'' ratings.

\begin{figure}[htbp]
    \centering
    \begin{minipage}[c]{0.50\columnwidth}
        \centering
        \includegraphics[width=\columnwidth]{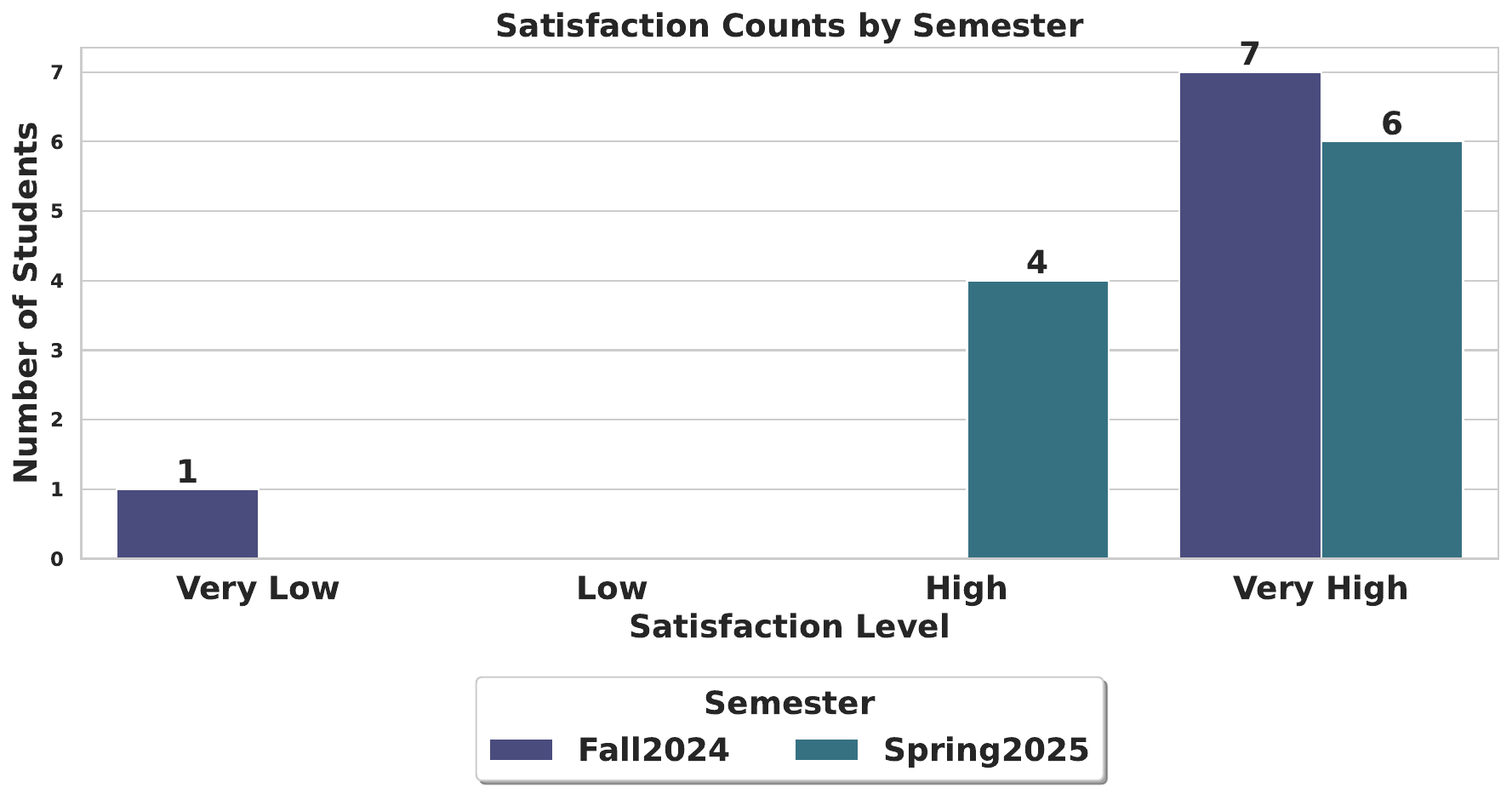}
        \caption{Bar Plot of Satisfaction Counts by Semester.}
        \label{fig:figure1_satisfaction_counts}
    \end{minipage}\hfill
    \begin{minipage}[c]{0.50\columnwidth}
        \centering
        \includegraphics[width=\columnwidth]{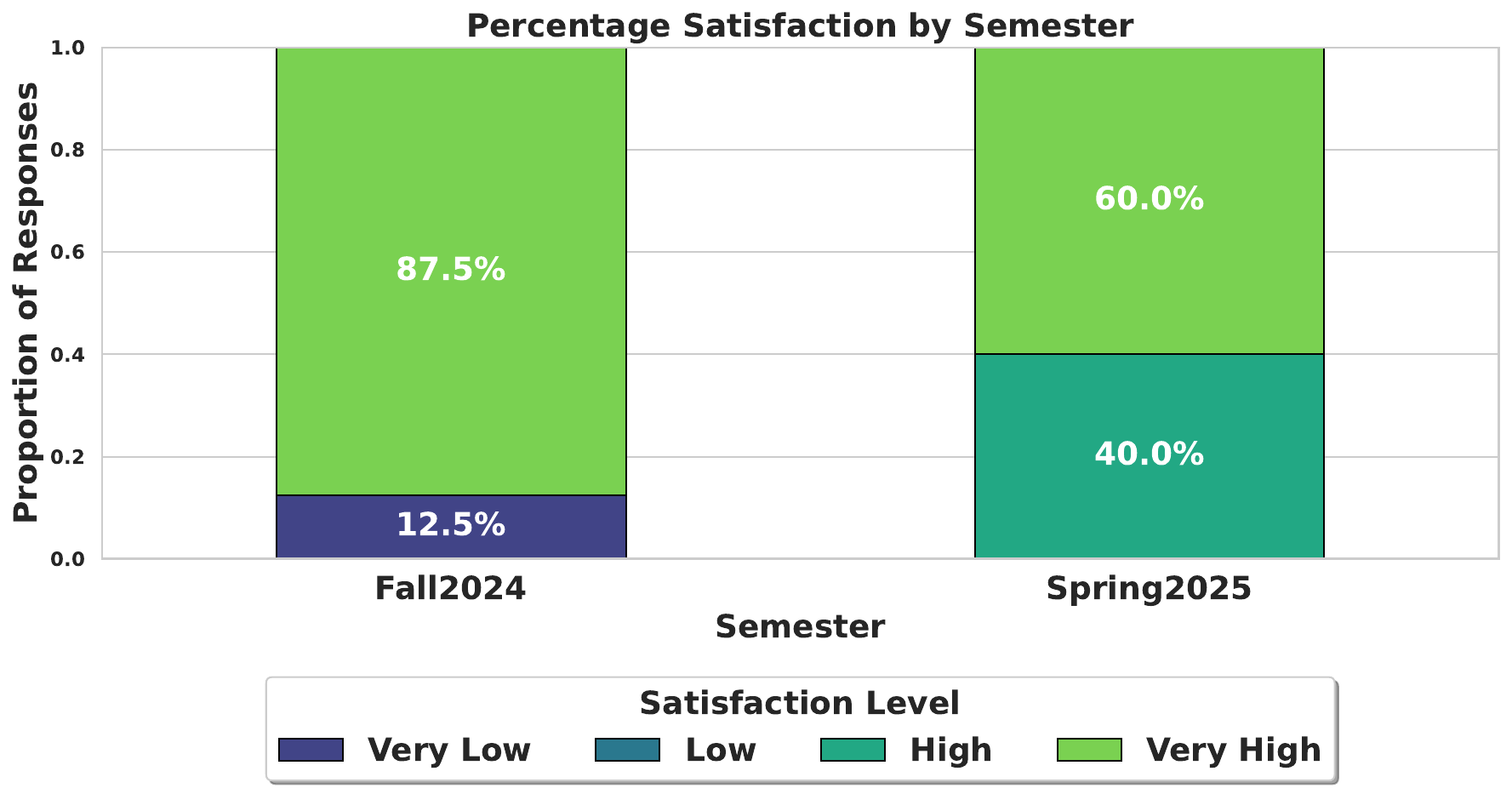}
        \caption{Stacked Bar Chart of Percentage Satisfaction by Semester.}
        \label{fig:figure2_percentage_satisfaction}
    \end{minipage}
\end{figure}

As shown in Figure~\ref{fig:figure1_satisfaction_counts} we counted the number of responses in each satisfaction category by semester. When we examined the proportional breakdown in Figure~\ref{fig:figure2_percentage_satisfaction} we saw that Fall 2024's responses clustered heavily around ``Very High,'' whereas Spring 2025 had a more balanced split between ``Very High'' and ``High.'' 


Overall, we conclude that students were very satisfied with the GPU programming course in both semesters. While Fall 2024 included an isolated low satisfaction rating, Spring 2025 showed consistently positive and moderately varied satisfaction levels. These findings suggest that the GPU programming course effectively facilitated the acquisition of valuable new skills, which contributed to high levels of student satisfaction, reflecting the course's successful delivery and instructional quality.

\end{document}